\documentclass{article}[12pt]

\begin{document}
{\noindent \it{\small ---}}
\begin{center}
\vskip 3.0cm
{\Large \bf Avalanches of popping bubbles \\
in collapsing foams}
\vskip 1.5cm
{\large \bf N.Vandewalle$^{1,2}$, J.F.Lentz$^1$, S.Dorbolo$^1$ and
F.Brisbois$^1$}
\vskip 0.5cm
$^1$ GRASP, Institut de Physique B5, Universit\'e de Li\`ege, \\
B-4000 Li\`ege, Belgium.
\vskip 0.5cm
$^2$ Laboratoire des Milieux D\'esordonn\'es et H\'et\'erog\`enes, Tour 13, Case 86, \\
4 place Jussieu, F-75252 Paris Cedex 05, France.

\end{center}

\vskip 4.0cm
keywords: avalanches, foams, topological rearrangements
\newpage

{\bf Cellular structures are very common in nature \cite{review}.
Each cell of the cellular structure can be a bubble in a beer, a biological cell in a tissue \cite{tissue}, a grain in a polycrystal \cite{poly} or a magnetic domain in a solid \cite{magn}. Foams have becomed paradigms of disordered cellular systems. Among the physical properties of interest, one is the long-term behaviour of a froth driven by topological rearrangements. In the present work, we report acoustic experiments on foam systems. We have recorded the sound emitted by crackling cells during the collapsing of foams. The sound pattern is then analyzed using classical methods of statistical physics. Fundamental processes at the surface of the collapsing foam are found. In particular, size is not a relevant parameter for exploding bubbles.}

\vskip 1.0cm

Foams have been created by blowing air at the base of a water/soap
mixture in a cylindric vessel. A typical resulting foam is illustrated in Figure 1. Polygonal bubbles are observed near the air/foam interface while spherical bubbles are seen at the water/foam interface. The thickness of the foam layer can be controlled such that the foam can be considered as dry in the region of our interests, i.e. the air/foam interface. Near this interface, the evolution of dry foam is slow and driven by geometrical constraints like the motion of vertices and edges \cite{review}. In addition, many topological rearrangements such as the cell side switching or the vertex disappearance take also place in the foam \cite{review}. The combination of bubble area growth/decay and topological rearrangements induces a complex dynamics \cite{glazier} in which subtle correlations are found as e.g. described by the so-called Aboav-Weaire law \cite{aboav}.

Moreover, one should note that many bubble explosions occur at the air/foam interface. Indeed, each bubble at the surface presents a very thin and curved face which is more fragile than the planar faces located inside the foam. The explosion of the surface bubbles leads to a macroscopic collapse of the foam as usually observed in beer and soap froths. A relevant question concerns the type(s) of bubbles which is(are) exploding at the air/foam interface. Another fundamental question is whether the bubble explosions are correlated or not? If yes, what is the nature of these correlations?

In order to study the explosion of bubbles, a microphone has been placed
above the foam. The sound of popping cells has been recorded at a
sampling rate of 32 $kHz$. In order to minimize the external noise as well as any secondary reflection of acoustic waves, the setup has been placed in a special ``anechoic chamber". Different commercial soaps have been tested. Containers of various sections have been also used.

A typical recording of acoustic activity $A(t)$ (dimensionless) is presented in Figure 2. Peaks of various amplitudes are distributed along the time series. Each Peak represents the explosion of one bubble at the surface of the collapsing foam. The characteristic duration of a peak is typically $\tau_0 = 5 \mu s$. A white noise component having a very small amplitude is also present. In order to extract the exact position and size of each peak, we have numerically treated the time series. The noise has been first removed by selecting a lower cutoff for peak amplitudes (noise thresholding). The resulting filtered signal $\tilde{A}(t)$ is also shown in Figure 2. On time series which are 20 minutes long, we observe typically 5000 events. One should note that the exploding events are not homogeneously distributed along the time axis. Bursts of acoustic activity separated by periods of stasis are observed. This heterogeneous acoustic activity will be characterized herebelow.

After noise filtering, the power $P$ dissipated within the small interval $\Delta t \approx \tau_0$ is then calculated for each peak, i.e.
\begin{equation}
P = \int_{t_0-\Delta t /2}^{t_0+\Delta t /2} \tilde{A}^2(t) dt.
\end{equation} The dissipated power $P$ is given in arbitrary units;
nevertheless it is assumed to be proportional to the energy dissipated
during the explosion of the bubble membrane, i.e. to be proportional to the
surface area of the disappearing cell. Figure 3 presents a typical
histogram $h(P)$ of the frequency of peak occurence as a function of the
peak intensity $P$. This distribution presents a maximum and a long
``tail". The inset of Figure 3 shows a log-log plot of the long tail. For large $P$ values, $h(P)$ behaves like a power law
\begin{equation}
h(P) \sim P^{-\nu}.
\end{equation} This power law behaviour of the tail holds over 1.5 decades for best cases. Both the maximum position of $h(P)$ and the tail exponent $\nu$ have non-universal values i.e. depend mainly on the nature of the soap/water mixture. Exponent $\nu$ values ranging from 1.5 up to 3.0 have been found. The asymptotic power law behaviour indicates that the distribution of peak amplitudes is quite broad. In other words, no sharp cutoff is observed in $h(P)$. This implies that a wide variety of bubble sizes is exploding and that large bubbles are sometimes stronger than small ones. This experimental result is in contrast with the widely accepted and intuitive argument that only large bubbles are fragile and explode at the air/foam interface. The latter intuitive argument would however lead to a sharp exponential cutoff in $h(P)$. On the contrary, our experiment demonstrates that no critical bubble size exists. The stabilization of some large bubbles with respect to apparently weak smaller bubbles should find some explanation by considering the neighbouring environment of each bubble \cite{aboav2}. This should be confirmed by direct observation of a collapsing foam which is outside the scope of this work. 

A wide variety of bubbles (small and large ones) participates thus to the collapsing of the foam. Let us investigate whether correlations exist or not. Figure 4 presents in a log-log plot the histogram $h(\tau)$ of the interpeak durations $\tau$, i.e. the time intervals $\tau$ separating successive bubble explosions. Four different series obtained with four different soap/water mixtures in different containers are illustrated. For each analyzed series, the data points have been rescaled by a certain factor in order to emphasize that all series seem to behave like a power law
\begin{equation}
h(\tau) \sim \tau^{-\alpha}
\end{equation} with an exponent $\alpha = 1.0 \pm 0.1$. A power law with $\alpha=1$ is illustrated by the continuous line in Figure 4. Each histogram occupies about 2 decades. The value of the exponent $\alpha$ do not change when other types of soap/water mixtures are used. Only the total number of explosions as well as the total dissipated power may change. 

For a homogeneous (random) occurence of bubble explosions, one expects a poissonian distribution, i.e. an exponential decay of $h(\tau)$! The power law behaviour indicates that bubbles explosions are correlated events! Moreover, an unique value for $\alpha$ implies that the temporal correlations in the collapsing foam are universal. 

The power law behaviour of $h(\tau)$ and the longtail of $h(P)$ suggest that the energy release is discontinuous and quite similar to self-organized critical systems \cite{bakbook}. Simulations \cite{kawasaki} and experiments \cite{gopal} have indeed shown that a slowly driven foam can be described by avalanches having a broad distribution of event rate versus the energy release. In these studies, events are abrupt topological rearrangements (mainly the coarsening and the vanishing bubbles) while in the present study, events are exploding bubbles at the surface of a collapsing foam. One understands that the
explosion of a bubble implies a topological change for neighbouring bubbles which may propagate in the bulk of the foam as well as along the air/foam interface. These topological changes may destabilize other bubbles at the interface and thus create avalanches of popping bubbles. 

In summary, our acoustic experiments have put into evidence the intermittent and correlated character of popping bubbles in a collapsing dry foam. In other words, the dynamics of a collapsing foam is discontinuous and evolves by sudden bursts of activity separated by periods of stasis. These avalanches are correlated. Moreover, we have discovered that a wide variety of bubble sizes participate to the phenomenon.

\vskip 0.6cm
{\noindent \bf Acknowledgements}

NV and SD are financially supported by FNRS and FRIA respectively. Thanks to the CEDIA laboratory at the University of Li\`ege and in particular A.Calderon who provided an access to the anechoic chambers allowing for high quality acoustic recordings.

\vskip 0.5cm
Correspondance and requests should be addressed to Nicolas Vandewalle, e-mail: nvandewalle@ulg.ac.be

\newpage
{\noindent \large Figure Captions}

\vskip 0.5cm
{\bf Figure 1} -- Picture of a typical foam obtained by blowing air at the base of a water/soap mixture.

\vskip 0.5cm
{\bf Figure 2} -- Acoustic recording of crackling bubbles in a collapsing dry foam: (top) typical recording and (bottom) filtered recording for which a white noise component has been removed.

\vskip 0.5cm
{\bf Figure 3} -- The histogram $h$ of the power $P$ dissipated during each bubble explosion. The inset presents the tail of $h$ in a log-log plot. The continuous line represent a power law fit. 

\vskip 0.5cm
{\bf Figure 4} -- Log-log plot of the histogram $h$ for the interpeaks
durations $\tau$. Two cases are illustrated. The continuous line is a power law with an exponent $\alpha = 1$.

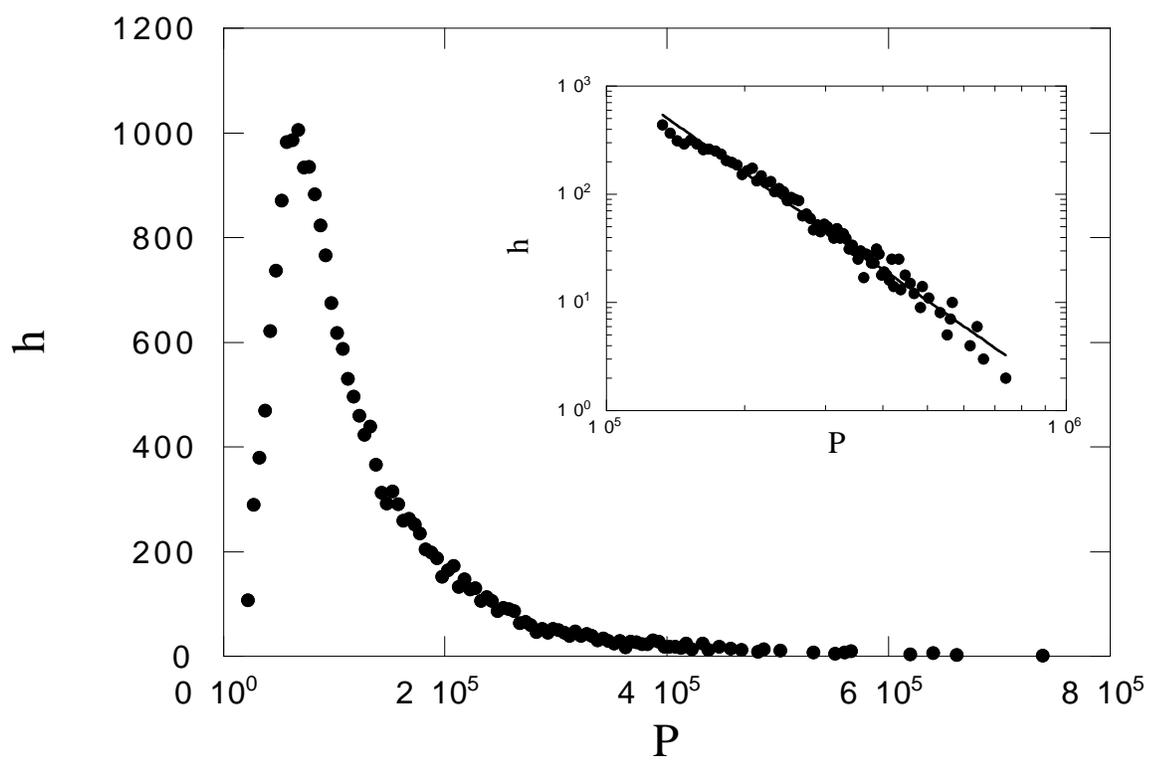
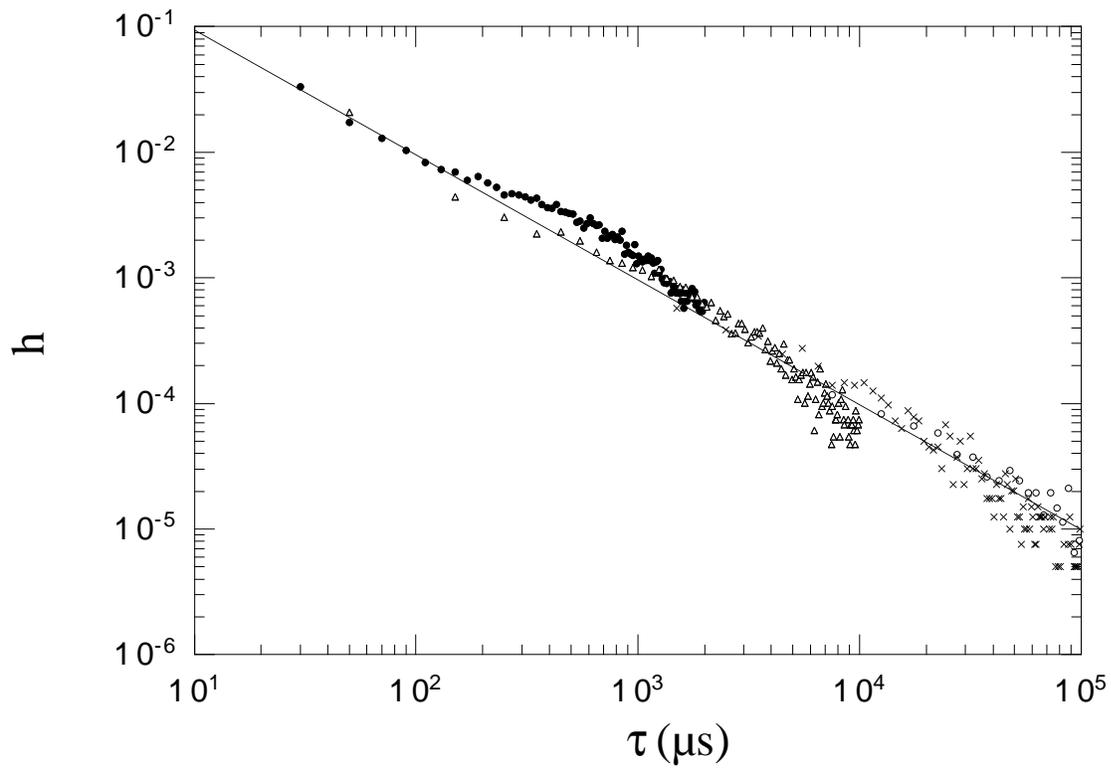
\end{document}